\documentclass{emulateapj}
\shorttitle{The Role of Continuum-Driven Eruptions}
\shortauthors{Smith \& Owocki}
\begin{document}

\title{ON THE ROLE OF CONTINUUM-DRIVEN ERUPTIONS IN THE EVOLUTION OF
  VERY MASSIVE STARS AND POPULATION III STARS}

\author{Nathan Smith\altaffilmark{1}}
\affil{Center for Astrophysics and Space Astronomy, University of
Colorado, 389 UCB, Boulder, CO 80309}

\author{Stanley P.\ Owocki}
\affil{Bartol Research Institute, University of Delaware, Newark, DE 19716}

\altaffiltext{1}{Hubble Fellow; nathans@casa.colorado.edu}

\begin{abstract}

We suggest that the mass lost during the evolution of very massive
stars may be dominated by optically thick, continuum-driven outbursts
or explosions, instead of by steady line-driven winds.  In order for a
massive star to become a Wolf-Rayet star, it must shed its hydrogen
envelope, but new estimates of the effects of clumping in winds from
O-type stars indicate that line driving is vastly insufficient.  We
discuss massive stars above roughly 40--50 M$_{\odot}$, which do not
become red supergiants, and for which the best alternative is mass
loss during brief eruptions of luminous blue variables (LBVs).  Our
clearest example of this phenomenon is the 19th century outburst of
$\eta$ Carinae, when the star shed 12--20 M$_{\odot}$ or more in less
than a decade.  Other examples are circumstellar nebulae of LBVs and
LBV candidates, extragalactic $\eta$ Car analogs (the so-called
``supernova impostors''), and massive shells around supernovae and
gamma-ray bursters.  We do not yet fully understand what triggers LBV
outbursts or what supplies their energy, but they occur nonetheless,
and present a fundamental mystery in stellar astrophysics.  Since line
opacity from metals becomes too saturated, the extreme mass loss
probably arises from a continuum-driven wind or a hydrodynamic
explosion, both of which are insensitive to metallicity.  As such,
eruptive mass loss could have played a pivotal role in the evolution
and ultimate fate of massive metal-poor stars in the early universe.
If they occur in these Population III stars, such eruptions would also
profoundly affect the chemical yield and types of remnants from early
supernovae and hypernovae thought to be the origin of long gamma ray
bursts.

\end{abstract}


\section{INTRODUCTION}

Mass loss is a critical factor in the evolution of a massive star.  In
addition to the direct reduction of a star's mass, it profoundly
affects the size of its convective core, its core temperature, its
angular momentum evolution, its luminosity as a function of time, and
hence its evolutionary track on the HR diagram and its main-sequence
(MS) lifetime (e.g., Chiosi \& Maeder 1986).  Wolf-Rayet (WR) stars
are the descendants of massive stars as a consequence of mass loss in
the preceding H-burning phases, during which the star sheds its H
envelope (Abbott \& Conti 1987; Crowther 2006).  While the maximum
initial mass of stars is thought to be $\sim$150 M$_{\odot}$ (Figer
2005; Kroupa 2005), WR stars do not have masses much in excess of 20
M$_{\odot}$ (Crowther 2006).\footnote{By ``WR stars'' we mean
H-deficient WR stars (core-He burning phases or later), and not the
luminous H-rich WNL stars (Crowther et al.\ 1995), which are probably
still core-H burning.}  Thus, very massive stars have the immense
burden of removing 30--130 M$_{\odot}$ during their lifetime before
the WR phase, unless they explode first.  Stellar evolution
calculations prescribe $\dot{M}$($t$) based on semiempirical values,
so it is important to know when most of this mass loss occurs.

In this letter we address the question of whether this mass loss
occurs primarily via steady stellar winds, or instead through violent,
short-duration eruptions or explosions.  Recent studies of hot star
winds indicate that mass-loss rates on the MS are much lower than
previously thought.  These mass-loss rate reductions are significant
enough to affect MS evolution, but they also raise an important
question: {\it If mass loss via stellar winds is insufficient to strip
off a star's H envelope and form a WR star, then how and when does it
occur}?  Simultaneously, observations of nebulae around luminous blue
variables (LBVs) and LBV candidates have revealed very high ejecta
masses -- of order 10 M$_{\odot}$.  In $\eta$ Car we know that the
mass was ejected in a single outburst and is not swept-up ambient
material.  Together, these facts suggest that short-duration outbursts
like the 19th century eruption of $\eta$ Car could dominate mass lost
during the lives of the most massive stars, and would be critical to
form WR stars.

As detailed below, the extreme mass-loss rates of these bursts imply
that line opacity is too saturated to drive them, so they must instead
be either continuum-driven super-Eddington winds or outright
hydrodynamic explosions.  Unlike steady winds driven by lines, the
driving in these eruptions may be largely independent of metallicity,
and might play a role in the mass loss of massive metal-poor stars
(Population III stars).

\section{THE PROBLEM: LINE-DRIVEN WINDS PROVIDE INSUFFICIENT MASS LOSS}

In order to shed a massive star's envelope and reach the WR stage,
models must prescribe semiempirical mass-loss rates, which can be
scaled by a star's metallicity (e.g., Chiosi \& Maeder 1986; Maeder \&
Meynet 1994; Meynet et al.\ 1994; Langer et al.\ 1994; Langer 1997;
Heger et al.\ 2003).  Often-adopted ``standard'' mass-loss rates are
given by de Jager et al.\ (1988), Nieuwenhuijzen \& de Jager (1990),
and Schaller et al.\ (1992).  In order for stellar evolution models to
match observed properties at the end of H burning, such as WR masses
and luminosities, and the relative numbers of WR and OB stars, these
mass-loss rates need to be enhanced by factors of $\sim$2 (Maeder \&
Meynet 1994; Meynet et al.\ 1994).

However, such enhanced mass-loss rates contradict observations.
Recent studies suggest that mass-loss rates are in fact 3--10 or more
times {\it lower} than the ``standard'' mass-loss rates, not higher.
Mass-loss rates based on density-squared diagnostics like H$\alpha$
and free-free radio continuum emission lead to overestimates if the
wind is strongly clumped.  Significant clumping in stellar winds is
expected based on theoretical considerations (Feldmeier 1995; Owocki
et al.\ 1988; Owocki \& Puls 1999), as well as observations like
time-variable discrete absorption components (Howarth et al.\ 1995;
Massa et al.\ 1995).  Recent efforts have thus focused on using
diagnostics that scale linearly with density, such as UV resonance
absorption lines; Fullerton et al.\ (2006) have suggested a reduction
of 10--20 or more from traditional mass-loss rates, while Bouret et
al.\ (2005) require reductions by factors of 3 or more (see also Puls
et al.\ 2006; Crowther et al.\ 2002; Hillier et al.\ 2003; Massa et
al.\ 2003; Evans et al.\ 2004).\footnote{Puls et al.\ (2006) express
concerns in UV-derived rates because of wind ionization, and
the reliability of tracers like P~{\sc v}.}  In any case, large
reductions in $\dot{M}$ are also needed to match the unexpectedly
symmetric X-ray line profiles in hot supergiant stars (Kramer et al.\
2003).\footnote{Suggestions (Oskinova et al.\ 2004) that the effective
reduction in bound-free X-ray absorption might instead be attributed
to a porous absorbing medium may require a separation scale between
clumps that is too large (Owocki \& Cohen 2006).}

Such reduced mass-loss rates mean that steady winds are simply
inadequate for the envelope shedding needed to form a WR star.  This
is not such a problem for stars below 10$^{5.8}$ L$_{\odot}$, where
the red supergiant (RSG) wind may be sufficient.  However, above
10$^{5.8}$ L$_{\odot}$ (initial mass above 40--50 $M_{\odot}$) stars
do not become RSGs (Humphreys \& Davidson 1979), posing a severe
problem if these stars depend upon line-driven winds for mass loss.

For example, consider the fate of a star with initial mass of 120
M$_{\odot}$.  The most extreme O2 If* supergiant HD93129A has a
mass-loss rate derived assuming a homogeneous wind of roughly
2$\times$10$^{-5}$ M$_{\odot}$ yr$^{-1}$ (Repolust et al.\ 2004).  If
the true mass-loss rate is lower by a factor of 3--10 or more as
indicated by clumping in the wind, then during a $\sim$2.5 Myr MS
lifetime (Maeder \& Meynet 1994), the star will only shed about 5--20
M$_{\odot}$, leaving it with M$\ga$100~M$_{\odot}$, and an additional
80~M$_{\odot}$ deficit to shake off before becoming a WR star.  After
this, the stellar wind mass-loss rates are higher during post-MS
phases, but they are still insufficient to form a WR star.  They
therefore cannot make up for the lower $\dot{M}$ values on the MS.
For a typical LBV lifetime of a few 10$^4$--10$^5$ yr (Humphreys 1989;
Bohannan 1997) and a typical $\dot{M}$ of $\sim$10$^{-4}$ M$_{\odot}$
yr$^{-1}$ for most LBVs, the LBV phase will only shed a few additional
solar masses through its line-driven wind.  Thus, some mechanism other
than just a steady wind is needed to reduce the star's total mass by
several dozen M$_{\odot}$.\footnote{Caveat: An important question
surrounds the lifetimes and evolutionary status of rare WNL stars
(e.g., Crowther et al.\ 2005), which have strong winds and may temper
the burder placed on LBVs if the WNL phase lasts $\ga$10$^6$
yr. However, in that case $\eta$ Car could not be a post-WNL star as
one might expect, because the mass of its ejecta added to its present
day stellar mass leave no room for such substantial mass loss if there
is an upper limit of 150 $M_{\odot}$ to the initial masses of stars
(Figer 2005; Kroups 2005).}

One obvious -- if not wildly speculative -- way out would be if WR
stars are {\it not} the decendants of the most massive stars because
they explode at the end of the LBV phase.  This, however, would be an
even more severe paradigm shift in our understanding of stellar
evolution, because it would require that LBVs have already reached
advanced core burning stages.  Even if that were the case, our central
hypothesis that continuum-driven LBV outbursts dominate the
pre-supernova mass loss would still be true because of the substantial
mass lost in LBV eruptions.

\section{AN ALTERNATIVE: LBV ERUPTIONS}

The most likely mechanism to rectify this hefty mass deficit is giant
eruptions of LBVs (e.g., Davidson 1989; Humphreys, Davidson, \& Smith
1999), where the mass-loss rate and bolometric luminosity of the star
increase substantially.  While we do not yet fully understand what
causes these giant LBV outbursts, we know empirically that they do
indeed occur, and that they drive substantial mass loss from the
star.

Our best example of this phenomenon is the 19th century ``Great
Eruption'' of $\eta$ Carinae.  The event was observed visually, the
mass of the resulting nebula has been measured (12--20 M$_{\odot}$ or
more; Smith et al.\ 2003b), and proper motion measurements of the
expanding nebula indicate that it was ejected in the 19th century
event (e.g., Morse et al.\ 2001).  The other example for which this is
true is the 1600 {\sc ad} eruption of P Cygni, although its shell
nebula has a much lower mass of only $\sim$0.1 M$_{\odot}$ (Smith \&
Hartigan 2006).  Both $\eta$~Car and P~Cyg are surrounded by multiple,
nested shells indicating previous outbursts (e.g., Walborn 1976;
Meaburn 2001).  While the shell of P~Cyg is less massive than
$\eta$~Car's nebula, it is still evident that P~Cyg shed more mass in
such bursts than via its stellar wind in the time between them (Smith
\& Hartigan 2006).  This difference between P~Cyg and $\eta$ Car hints
that LBV outbursts become progressively more extreme near the
Eddington limit.

For other LBVs surrounded by nebulae, we can't be certain that the
observed shells result from a single outburst, free of swept-up
stellar wind (e.g., Robberto et al.\ 1993).  However, upon comparison
with $\eta$ Car, it seems plausible that the observed range of nebular
masses originated in giant eruptions.  Deduced masses of LBV and
LBV-candidate nebulae from the literature are plotted in Figure 1 as a
function of the central star's luminosity.  We see that for stars with
log(L/L$_{\odot}$)$\ga$6, nebular masses of 10 M$_{\odot}$ are quite
reasonable, {\it perhaps suggesting that this is a typical mass
ejected in a giant LBV eruption}.

Figure 1 does not recover the clean ``nebular mass-stellar
luminosity'' relation of Hutsemekers (1994), which was based on just 6
objects.  In hindsight, we should not expect such a clean relation,
because it would indicate that a star of a given luminosity can only
eject a nebula of a particular mass.  In the case of $\eta$ Car we
know this is false: it ejected the very massive Homunculus in the
1840's, it ejected the 0.1--0.2 M$_{\odot}$ ``Little Homunculus'' in
1890 (Smith 2005; Ishibashi et al.\ 2003), and it may suffer smaller
ejections every 5.5 years (Davidson 1999; Smith et al.\ 2003a; Martin
et al.\ 2006).  Instead, we might expect a luminosity-dependent upper
threshold to the plot, populated underneath by a range of masses.

Although LBV eruptions are rare, a number of extragalactic $\eta$ Car
analogs or ``supernova impostors'' have been observed, such as SN1954J
in NGC2403 and SN1961V in NGC1058 (Humphreys et al.\ 1999; Smith et
al.\ 2001; Van Dyk et al.\ 2002, 2005), V1 in NGC2363 (Drissen et al.\
1997), and several recent events seen as type IIn supernovae, like
SN1997bs, SN2000ch, SN2002kg, and SN2003gm (Van Dyk et al.\ 2000,
2006; Wagner et al.\ 2004; Weis \& Bomans 2005; Maund et al.\ 2006).
Furthermore, massive circumstellar shells have also been inferred to
exist around supernovae and gamma-ray bursters (GRBs).  Some examples
are the radio-bright SN1988Z with a nebula as massive as 15
M$_{\odot}$ (Aretxaga et al.\ 1999; Williams et al.\ 2002; Van Dyk et
al.\ 1993; Chugai \& Danziger 1994), as well as similar dense shells
around SN2001em (Chugai \& Chevalier 2006), SN1994W (Chugai et al.\
2004), SN1998S (Gerardy et al.\ 2002), GRB021004 (Mirabal et
al. 2003), and GRB050505 (Berger et al.\ 2005).

These outbursts and the existence of massive circumstellar nebulae
indicate that the 19th century eruption of $\eta$ Car is not an
isolated, freakish event, but instead may represent a common rite of
passage in the late evolution of the most massive stars. A massive
ejection event may even initiate the LBV phase, by lowering the star's
mass, raising its L/M ratio, and drawing it closer to instability
associated with an opacity-modified Eddington limit (Appenzeller 1986;
Davidson 1989; Lamers \& Fitzpatrick 1988).  Mass loss in these giant
eruptions may play a role in massive star evolution analogous to
thermal pulses of asymptotic giant branch stars.  In any case, meager
mass-loss rates through stellar winds, followed by huge bursts of mass
loss in violent eruptions at the end of core-H burning may
significantly alter stellar evolution models.

\begin{figure}
\epsscale{0.99}
\plotone{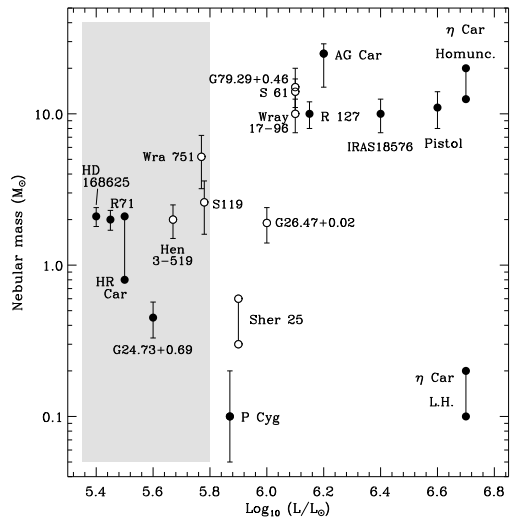}
\caption{Masses of ejecta nebulae from LBVs (filled dots) and LBV
  candidates (unfilled) as a function of the central star's bolometric
  luminosity.  Luminosities are taken from Smith, Vink, \& de Koter
  (2004), while masses are taken as follows: the Homunculus of $\eta$
  Car (Smith et al.\ 2003b), the Little Homunculus of $\eta$ Car
  (Smith 2005), the Pistol star (Figer et al.\ 1999), IRAS
  18576+0341/AFGL2298 (Ueta et al.\ 2001), AG Car and Wra 751 (Voors
  et al.\ 2000), G79.29+0.46 (Higgs et al.\ 1994), Wray 17-96 (Egan et
  al. 2002), Sher 25 (Brandner et al.\ 1997), P~Cygni (Smith \&
  Hartigan 2006), Hen 3-519 (Smith et al.\ 1994), and the remaining
  values adopted from Clark et al.\ (2003).  When masses are
  determined from measurements of dust masses, we assume a gas:dust
  mass ratio of 100.  When uncertainties are not specified by authors,
  we adopt roughly $\pm$25\%.  The lightly shaded part on the left
  side of the graph corresponds to luminosities of stars that may be
  post-red supergiants (see Smith, Vink, \& de Koter 2004).}
\end{figure}

\section{EXTREME MASS-LOSS RATES AND OPTICALLY THICK CONTINUUM-DRIVEN WINDS}

Observational constraints require extremely high mass-loss rates
during giant LBV eruptions.  For $\eta$ Car, we have a lower limit of
0.5~M$_{\odot}$ yr$^{-1}$ averaged over the 20 yr duration of the
eruption (Smith et al.\ 2003b).  However, the thin walls of the
Homunculus (Smith 2006) and the small age spread from proper motions
(Morse et al.\ 2001) both imply that the dominant mass-loss phase was
$\la$5 yr.  This would indicate an astonishing mass-loss rate of
several M$_{\odot}$ yr$^{-1}$ or more.  Furthermore, the 20 yr bright
phase of $\eta$ Car was unusually long-lasting; eruptions of
extragalactic $\eta$ Car-analogs typically last less than a decade
(Van Dyk 2005).  P Cygni presents the lower end of the spectrum for
likely mass-loss rates.  Its outburst in 1600 {\sc ad} ejected
$\sim$0.1 M$_{\odot}$ (Smith \& Hartigan 2006), implying
$\dot{M}\approx$10$^{-2}$ M$_{\odot}$ yr$^{-1}$.

Such extreme mass-loss rates mean that strong lines must be heavily
saturated, so that these outflows cannot be launched by the
conventional CAK (Castor et al.\ 1975) mechanism for line-driven
winds.  As discussed by Owocki et al.\ (2004), the maximum mass-loss
rate for line driving can be written as

\begin{equation}
{\dot M} = { L \over c^{2} } {\alpha \over 1-\alpha} 
\left [ { {\bar Q} \Gamma_{e} \over 1 - \Gamma_{e} } 
\right ]^{-1+1/\alpha}
\, ,
\label{mdcak}
\end{equation}

\noindent where $L$, $c$, and $\Gamma_{e}$ are the stellar luminosity,
speed of light, and Eddington parameter (for pure electron
scattering), and $\alpha$ and ${\bar Q}$ are the power index and
normalization of the line opacity distribution (Gayley 1995).  This
mass loss scaling arises from the need for the line acceleration --
which scales inversely with density and thus mass loss rate -- to
overcome gravity in driving the wind.  For stars close enough to the
Eddington limit that the effective gravity becomes small, the mass
loss can formally become large.  However, this would also result in
outflow speeds that are smaller than inferred from observations of
LBVs.  To characterize the maximum mass loss that can be driven
without this kind of augmentation from a separate continuum
assistance, let us take the factor $\Gamma_{e}/(1-\Gamma_{e})$ to be
roughly unity.  Then for optimal realistic values $\alpha=1/2$ and
${\bar Q} = 2000$ for the line opacity parameters (Gayley 1995), the
maximum mass loss from line driving is given by

\begin{equation}
{\dot M} \approx 1.4 \times 10^{-4} \, L_6 \, M_{\odot} \, yr^{-1} \, ,
\end{equation}

\noindent where $L_6$ is L/(10$^6$ L$_{\odot}$).  Even for peak
luminosities of few $10^{7} L_{\odot}$ during $\eta$ Car's eruption,
this limit is still several orders of magnitude below the mass-loss
that created the Homunculus.  If mass loss during these eruptions
occurs via a wind, it must be a super-Eddington wind driven by
continuum radiation pressure (i.e., Thomson scattering opacity and not
lines; Owocki et al.\ 2004; Belyanin 1999; Quinn \& Paczynski 1985).

An alternative to a continuum-driven wind is a deep seated
hydrodynamic explosion that blasts off the star's outer layers.  In
the star's envelope, convection will set in before the Eddington limit
is reached, but if convection is inefficient, a density inversion can
develop (e.g., Joss et al.\ 1973).  Potentially, this could lead to a
violent explosion (e.g., Arnett et al.\ 2005; Young 2005).
Gravity-mode oscillations or non-linear growth of other instabilities
within the star may also play a role (Glatzel et al.\ 1999; Townsend
\& MacDonald 2005; Guzik 2005).  It is not yet clear which of these
phenomena is responsible for giant LBV eruptions, but none of them
invokes metallicity-dependent line driving as the physical mechanism
for imparting momentum to the ejecta.

\section{POTENTIAL IMPLICATIONS FOR THE FIRST STARS}

The first stars, which should have been metal free, are generally
thought to have been predominantly massive, exhibiting a flatter
initial mass function than stars at the present epoch (e.g., Bromm \&
Larson 2004).  With no metals, these stars should not have been able
to launch line-driven winds, and thus, they are expected to have
suffered no mass loss during their lifetimes.  The lack of mass loss
profoundly affects the star's evolution and the type of supernova it
eventually produces (Heger et al.\ 2003), as well as the yield of
chemical elements that seeded the early interstellar medium of
galaxies.

This view rests upon the assumption that mass loss in massive stars at
the present time is dominated by line-driven winds -- an assumption
that is problematic in view of recent observational constraints.  As
discussed above, massive shells around LBVs and the so-called
``supernova impostors'' in other galaxies indicate that short-duration
eruptions dominate the mass loss of very massive stars, while steady,
line-driven winds contribute little to the total mass lost during
their lifetime.  Unlike line-driven winds, the driving mechanism for
these outbursts could well be insensitive to metallicity.

Since we still do not know what triggers LBV eruptions, we cannot yet
claim confidently that these eruptions will in fact occur in the first
stars.  However, the possibility should raise caution signs for
theoretical work on Population III stars.  If mass loss of massive
stars at the present epoch is dominated by mechanisms that are
insensitive to metallicity, then we must question the prevalent notion
that the first stars did not lose substantial mass prior to their
final supernova event.

\acknowledgments \scriptsize

We thank S.\ Van Dyk, A.\ Gal-Yam, D.\ Arnett, S.\ Smartt, J.\ Puls,
and P.\ Crowther for several interesting discussions.  N.S.\ was
supported by NASA through grant HF-01166.01A from STScI, which is
operated by the Association of Universities for Research in Astronomy,
Inc., under NASA contract NAS5-26555.  S.P.O.\ acknowledges support of
NSF grant AST-0507581.


\end{document}